\documentclass[cits]{PoS}
\usepackage{amssymb, fontenc, times, mathptmx}
\usepackage{amsmath,latexsym,amsfonts}
\newcommand{\mc}[1]{\ensuremath{\mathcal{ #1 }}}
\newcommand{\avg}[1]{\left< #1 \right>}
\newcommand{\p}{\ensuremath{\partial}}
\newcommand{\eq}[1]{\begin{equation} #1 \end{equation}}

\title{Charm physics with Moebius Domain Wall Fermions}

\ShortTitle{Charm physics with Moebius Domain Wall Fermions}

\author{Andreas J\"uttner, Francesco Sanfilippo,
  \speaker{Justus Tobias Tsang}\\
        School of Physics and Astronomy, University of Southampton\\
        SO17 1BJ Southampton, United Kingdom\\
        E-mail: \email{j.t.tsang@soton.ac.uk}}

\author{Peter Boyle,\\
        School of Physics and Astronomy, University of Edinburgh\\
        EH9 3JZ, Edinburgh, United Kingdom}

\author{{Marina Marinkovi\'c,}\\
        School of Physics and Astronomy, University of Southampton\\
        SO17 1BJ Southampton, United Kingdom and\\
        CERN, Physics Department, 1211 Geneva 23, Switzerland\\
}

\author{{Shoji Hashimoto, Takashi Kaneko}\\
  High Energy Accelerator Research Organization (KEK) and\\
  School of High Energy Accelerator Science, The Graduate University
  for Advanced Studies,\\
  Tsukuba 305-0801.
} 

\author{{Yong-Gwi Cho}\\
  Graduate School of Pure and Applied Sciences, University of Tsukuba,\\
  Tsukuba, Ibaraki 305-8571.
}

\abstract{We present results showing that Domain Wall fermions are a suitable discretisation for the simulation of heavy quarks. This is done by a continuum scaling study of charm quarks in a M\"obius Domain Wall formalism using a quenched set-up. We find that discretisation effects remain well controlled by the choice of Domain Wall parameters preparing the ground work for the ongoing dynamical $2+1f$ charm program of RBC/UKQCD.}

\FullConference{The 32nd International Symposium on Lattice Field Theory\\
		 23-28 June, 2014\\
		 Columbia University New York, NY}

\begin{document}

\section{Introduction}
To determine potential channels for New Physics (NP) it is necessary to test and over-constrain the Standard Model (SM) by performing model independent theoretical predictions with high precision. For this all systematic errors need to be controlled. Currently the heavy quark sector (charm and bottom) is rather unexplored compared to the light quark sector (see \cite{FLAG}) which makes it interesting for numerical simulations in Lattice Quantum Chromodynamics (LQCD). However, the simultaneous simulation of light and heavy (charm) quarks poses a difficulty. Namely, in order to control finite volume effects for the light quarks large lattice volumes are required, whilst to resolve the heavy quarks it is necessary to have fine lattice spacings $a$, which is very costly to achieve at the same time. The work presented in this talk investigates the feasibility of simulating charmed mesonic quantities within a Domain Wall formalism to lay the ground work for RBC/UKQCD's ongoing $2+1f$ charm program at the physical point whilst maintaining chiral symmetry and hence $O(a)$ improvement \cite{juettner:Lattice14}.

To this end we produced 4 tree-level improved Symanzik gauge ensembles with inverse lattice spacings $a^{-1}$ between $2\mathrm{GeV}$ and $5.6\mathrm{GeV}$ in an approximately constant volume allowing for a controlled continuum limit. We stress that this quenched set-up is not intended to make physical predictions, but to have a feasible framework to test the applicability of M\"obius Domain Wall fermions to simulations of heavy quark physics. This is done by investigating the scaling behaviour in the continuum limit approach. To reduce the cost and since the simulated volume is comparably small, no light quarks are simulated but instead we consider $\eta_s$-like $D_s$-like and $\eta_c$-like quantities. Using the quenched approximation allows the exploration of very fine lattice spacings which are not otherwise affordable. Furthermore we developed the expertise and understanding of the parameter space which is needed for the dynamical simulations and are now set-up for the aforementioned ongoing $2+1f$ measurements presented at this conference ~\cite{juettner:Lattice14}. A similar study comparing different valence discretisations for the heavy quarks was carried out on the same gauge ensembles and also presented at this conference \cite{Cho:Lattice14}.

\section{Ensembles}

\begin{table}
\begin{center}
\begin{tabular}{c c c c c c c c}
\hline\hline
$\beta$ & $L/a$   & $N_\textrm{HB}$  & $N_\textrm{sep}$ 	& $\tau_{int}(Q_\textrm{top})$ 	& $\tau(Q^2_\textrm{top})$  &  $a^{-1}[GeV]$ & $L[\mathrm{FD}$]\\\hline
4.41 	& 16  	  & 10000           & 100	   	& 15(3)                         & 10.5(1.6)                &2.00(07)       & 1.579(55)\\
4.66 	& 24   	  & 20000           & 200	        & 159(60)                       & 74(22)                   &2.81(09)       & 1.686(52)\\
4.89 	& 32   	  & 600000 	    & 500 	        & 215(100)	                & 167(80) 	           &3.80(12)       & 1.664(51)\\
5.20	& 48      & 1400000 	    & 40000 	   	& $139(63) \times 200$		& $58(21) \times 200$	   &5.64(22)       & 1.683(64)\\
\hline\hline 
\end{tabular}
\caption{Run parameters and autocorrelation times for the topological charge for the simulated tree-level-improved Symanzik gauge ensembles. The lattice volume is kept fixed to $L\approx1.6~\mathrm{fm}$. The separation between the saved configurations of $\sim~2\times \tau_{int} (Q^2)$ allows for the autocorrelation between the saved configurations to be neglected.}
\label{tab:runpars}
\end{center}
\end{table}

We perform all our measurements on four sets of $\mc{O}(100)$ independent, tree-level-improved Symanzik gauge configurations  \cite{Curci1983a,Luscher1985}, where the inverse lattice spacing is varied in the range $2-5.6~\mathrm{GeV}$ and the physical volume is kept fixed to $\sim1.6\mathrm{fm}$ (cf. table \ref{tab:runpars}). 
For the generation of SU(3) gauge configurations, we use the heat-bath (HB) algorithm implemented in CHROMA \cite{Edwards2005} as well as additional code optimised for BG/Q.
In Table \ref{tab:runpars} we give the parameters of the performed runs for the configuration generation and the autocorrelation times of the topological charge Q,
\begin{align}
Q=\frac{1}{16\pi^2} a^4 \sum_x   Tr \{ F_{\mu\nu}(x) \tilde{F}_{\mu\nu}(x)\},
\end{align}
where $F_{\mu\nu}$ represents a clover definition of the field strength tensor. The topological charge was closely monitored to ensure ergodic sampling. The square topological charge $Q^2$ is considered to be a slowest mode relevant for the dynamics of the simulated system \cite{crit_slow_down_slowest_mode} and we use its autocorrelation time (found as described in \cite{autocorrelationprocedure}) to estimate the minimal separation between the saved configurations, so that they can be considered independent. 

\subsection{Scale setting}

\begin{figure}
\begin{center}
\includegraphics[width=.47\textwidth]{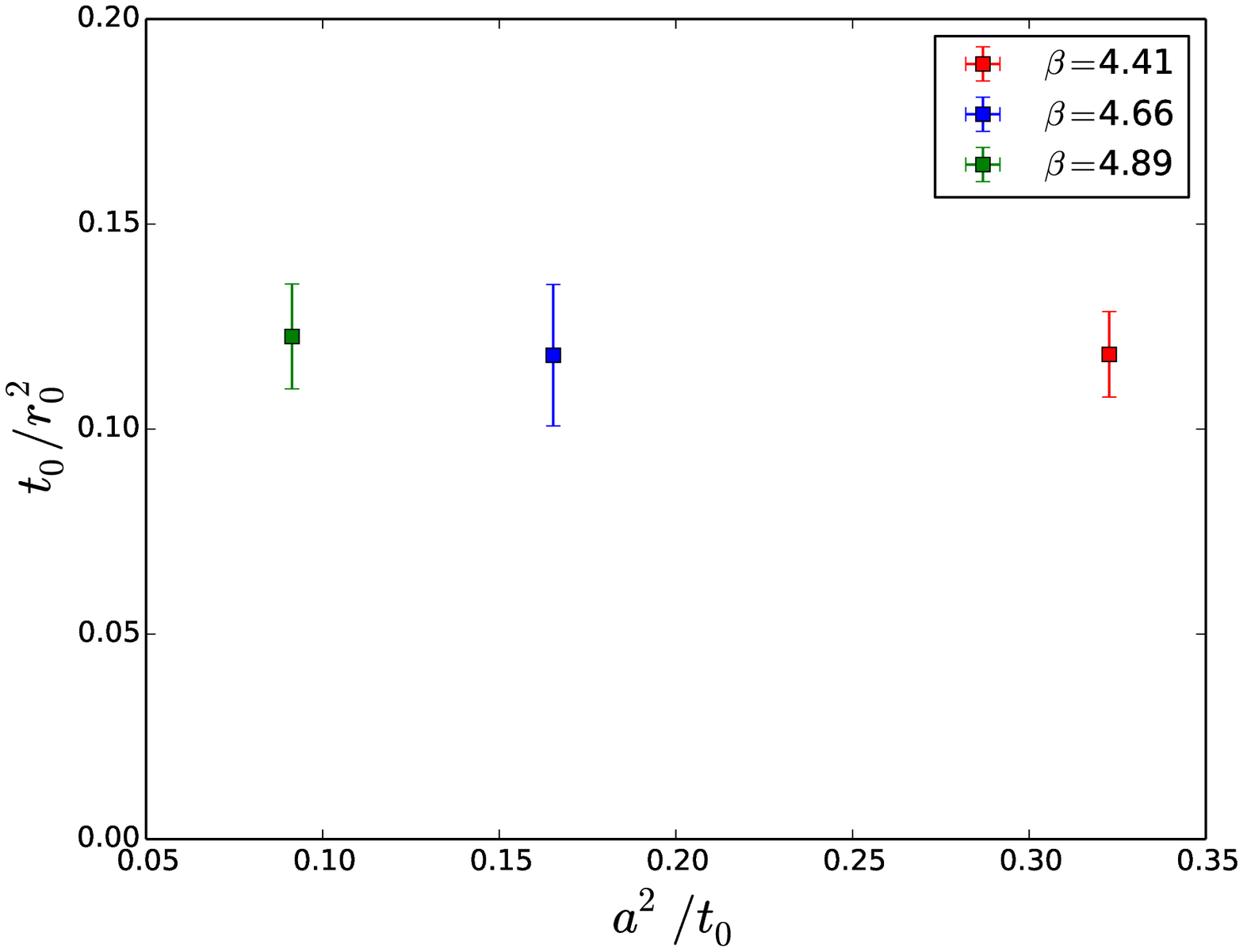}
\includegraphics[width=.45\textwidth]{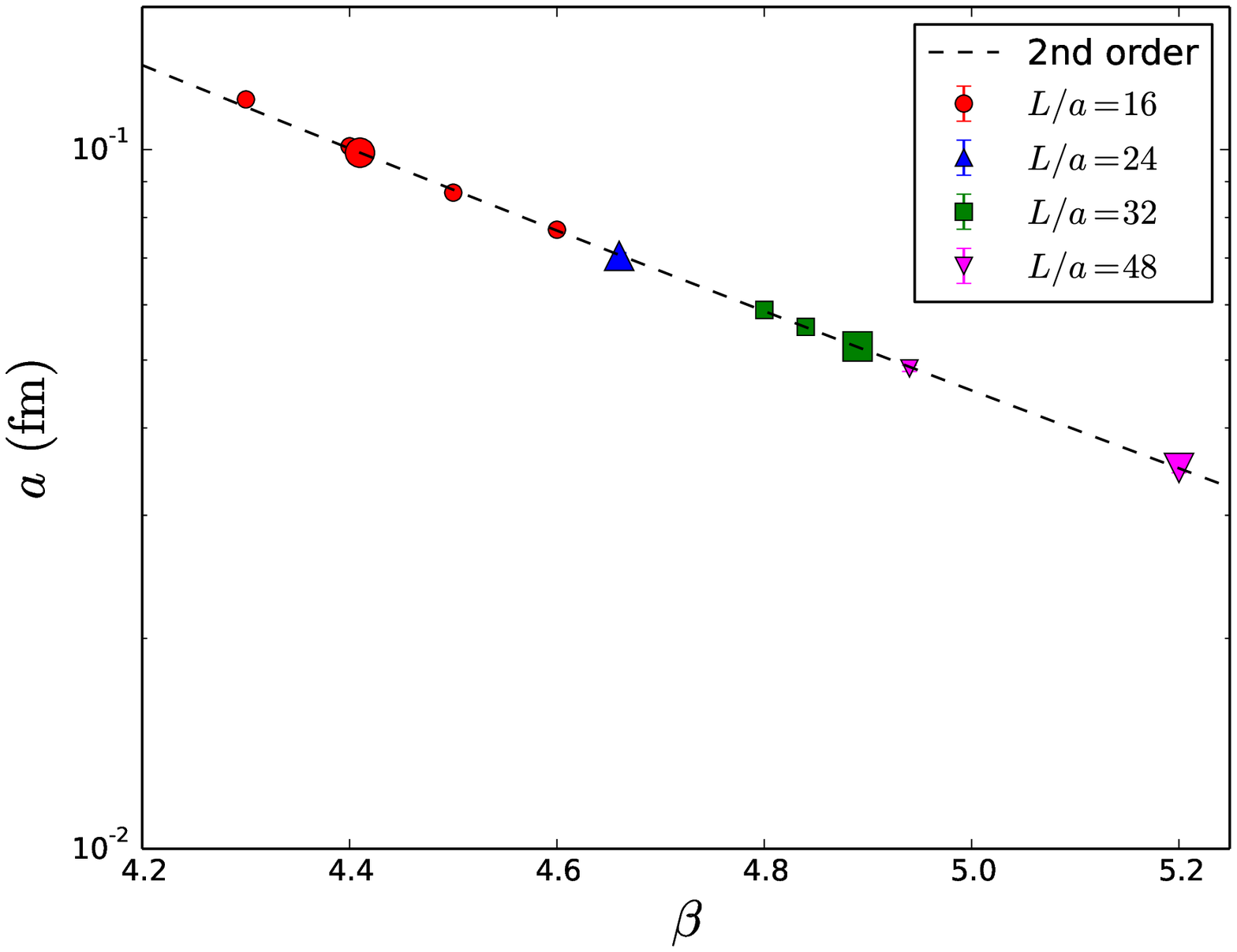}
\end{center}
\caption{Left: Determination of the lattice spacing with the Sommer scale parameter $r_0$ and flow parameter $t_0$ as a cross check of the determination of the lattice spacing. Right: $\log{a}$ as a function of $\beta$ to determine finite size effects. 
The larger symbols correspond to the 4 ensembles used for measurements, the smaller to the auxiliary ensembles with volumes of $1.23-1.89\mathrm{fm}$.}
\label{fig:scaling}
\end{figure}

We use the Wilson flow \cite{Luscher:2009eq} to set the scale of our simulations, in particular, we use the $w_0$ -~scale which was proposed in \cite{Borsanyi:2012zs}. From this we also take the physical scale $w_0^\mathrm{phys}=0.1755(18)(04)\mathrm{fm}$, computed for $2+1f$ simulations. The program package GLU\footnote{https://github.com/RJhudspith/GLU}{} used for the Wilson flow measurements was kindly provided by Jamie Hudspith.
We additionally 
perform a cross check of the scale setting procedure by measuring the Sommer scale parameter $r_0$ and an additional flow parameter $t_0$ ~\cite{Luscher:2010iy} on the 3 coarsest ensembles. It is found that these agree well (cf. l.h.s of figure \ref{fig:scaling}). The one-loop beta function was used to estimate the parameters of the simulations. From the r.h.s. of figure \ref{fig:scaling} it can be seen that all lattices with volumes larger than $\sim1.2\mathrm{fm}$ (i.e. all points other than the grey squares in figure \ref{fig:scaling}) are consistent with the fitted scaling curve. The universal scaling curve for all simulated volumes larger than $1.2\mathrm{fm}$ confirms the observation of ref. \cite{Borsanyi:2012zs} that the finite size effects of the particular definition of the Wilson flow parameter are negligible in the volume range we are interested in.


\section{Parameter Choices}
In this study we aim to determine the range of parameter space in which Domain Wall fermions (DWF) can be used to reliably simulate heavy quarks with O(a) improvement. The use of domain wall fermions introduces two new parameters which can be chosen freely: The Domain Wall height $M_5$ and the extent of the fifth dimension $L_s$.  In this study we use two different versions of the Domain Wall formalism both of which are used in the RBC/UKQCD $2+1f$ simulations, namely the 'standard' \emph{Shamir} Domain Wall fermions \cite{Kaplan:1992bt,Shamir:1993zy} and the more recently developed \emph{M\"obius} Domain Wall fermions \cite{Brower:2004xi,Brower:2005qw,Brower:2012vk}. M\"obius Domain Wall fermions are a rescaled version of the Shamir case, allowing to simulate the same physics with half the extent in the 5th dimension and therefore decreasing the computational cost. In particular choosing $L_s^{\text{M\"obius}} = L_s^\mathrm{Shamir}/2$ is expected to reproduce the same results.

\begin{figure}
\begin{center}
\includegraphics[width=.45\textwidth]{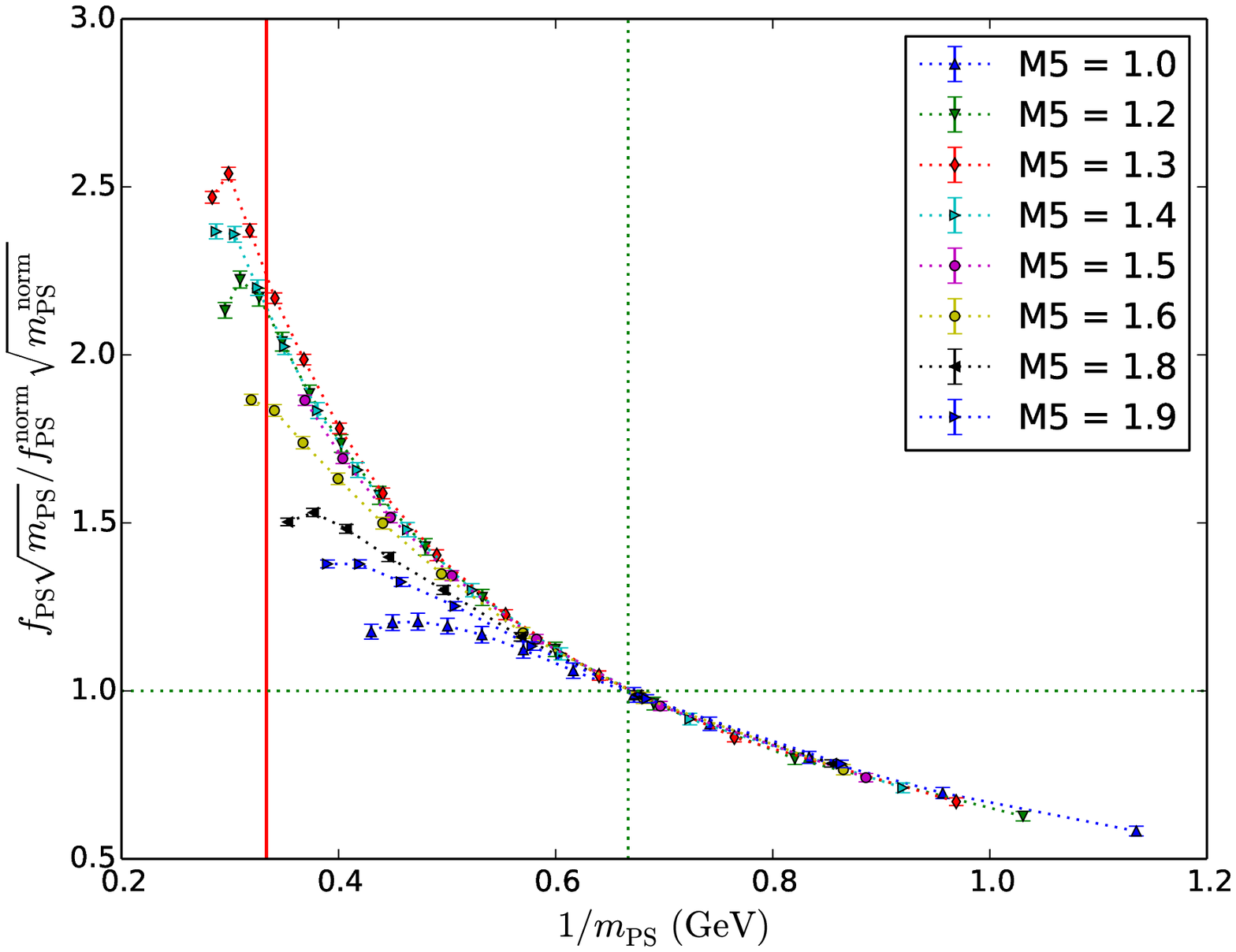}
\includegraphics[width=.45\textwidth]{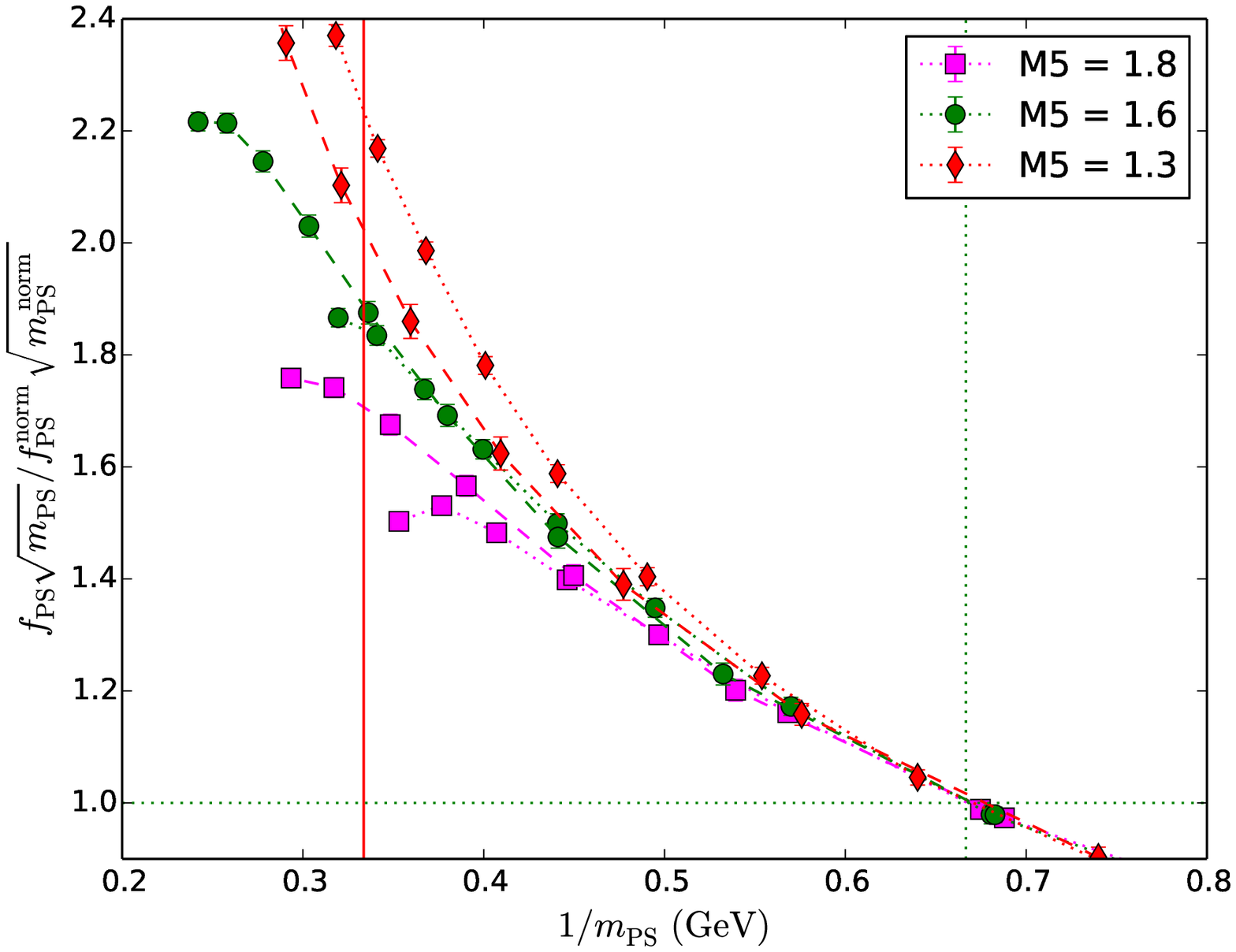}
\end{center}
\caption{Left: Behaviour of $f_{\mathrm{PS}}\sqrt{m_\mathrm{PS}}$ for heavy-heavy pseudo scalar mesons as a function of the inverse pseudo scalar mass $m_{\mathrm{PS}}$ for different values of $M_5$ on the coarsest ensemble using Shamir DWFs with $L_s=16$. The data is normalised at $m^\mathrm{norm}_\mathrm{PS} = 1.5\mathrm{GeV}$ (horizontal and vertical green dotted lines) to avoid finding a renormalisation constant. The vertical red line corresponds to the $\eta_c$ mass. Right: Overlay of data from the two coarsest ensembles for 3 choices of $M_5$. The dotted (dashed) curves depict data from the (second) coarsest ensemble.}
\label{fig:kink}
\end{figure}

When naively increasing the bare quark mass, unphysical behaviour is found for $am_q^\mathrm{bare} \gtrsim 0.4$. This can be seen from the unexpected bending down of $f_\mathrm{PS}\sqrt{m_\mathrm{PS}}$ as heavier pseudo scalar masses are approached (compare left hand panel of figure \ref{fig:kink}). The right hand panel of figure \ref{fig:kink} suggests that this behaviour is very sensitive to the choice of $M_5$. Additional to the pseudo scalar mass for which this bending sets in, the size of the cut-off effects also depends strongly on the choice of $M_5$. We found that the optimal choice of the Domain Wall height is given by $M_5=1.6$, minimising cut-off effects at the same time as postponing the unphysical behaviour to very close to charm even on the coarsest ensemble.
In the limit $L_s \to \infty$, Domain Wall fermions maintain exact chiral symmetry on the lattice. However for finite $L_s$ there is a residual chiral symmetry breaking that can be quantified by the \emph{residual mass} $m_{\mathrm{res}}$ which can be defined from the axial Ward Identity for the Domain Wall fermions
\eq{
\avg{\p_\mu \mc{A}_\mu(x)P(0)} = 2 m \avg{P(x) P(0)} + 2 \avg{J_{5q}(x)P(0)},
}
so that
\eq{
am_{\mathrm{res}} = \frac{\sum_x J_{5q}(x)P(0)}{\sum_x P(x)P(0)}.
\label{mres}
}
Here $\mc{A}_\mu$ is the conserved axial current, $P$ is the pseudo scalar density and $J_{5q}$ is a pseudo scalar current that mediates between the boundaries of the 5th dimensions and the centre of the 5th dimension.

\begin{figure}
\begin{center}
\includegraphics[width=.45\textwidth]{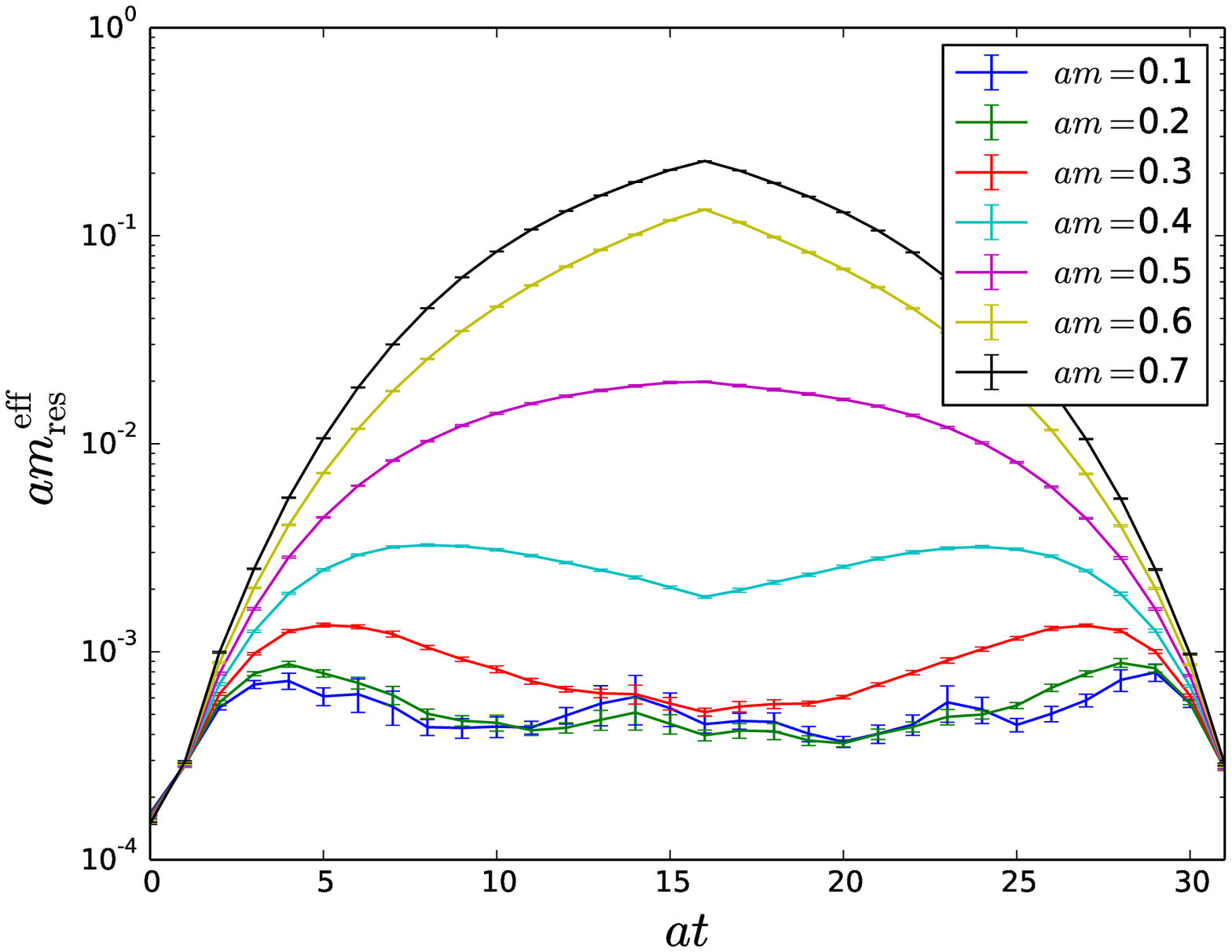}
\includegraphics[width=.45\textwidth]{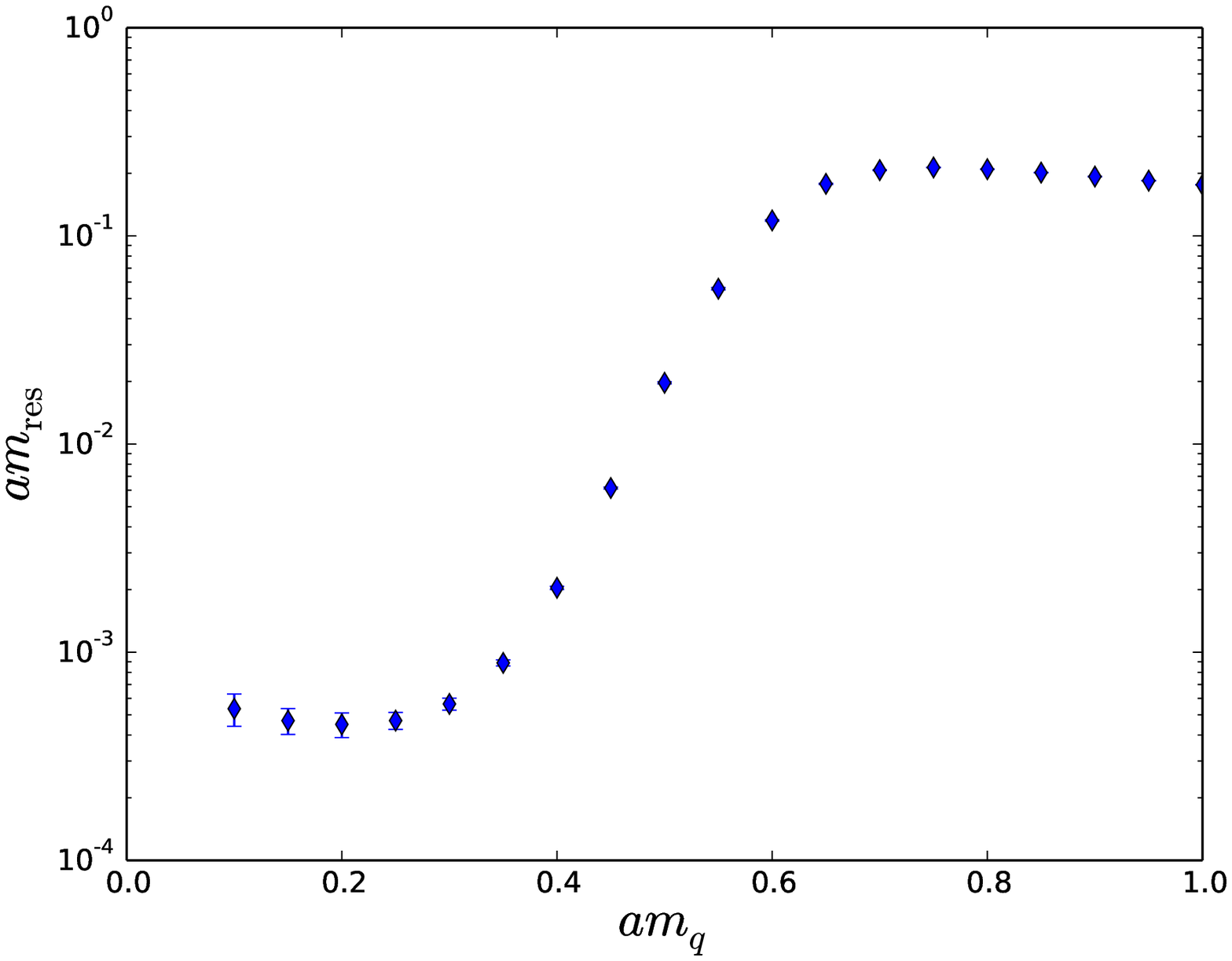}
\end{center}
\caption{Left: Behaviour of the \emph{effective} residual mass $m_{\mathrm{res}}^{\mathrm{eff}}$ as a function of time. $m_{\mathrm{res}}^{\mathrm{eff}}$ is found by evaluating (3.2) on every time slice. Right: The residual mass taken at the centre of the lattice as a function of the bare quark mass. Both panels are for the coarsest ensemble with Shamir Domain Wall fermions and $L_s=16$.}
\label{fig:mres}
\end{figure}

To investigate the above mentioned unphysical behaviour the residual mass was closely monitored (compare figure \ref{fig:mres}) and from this it can be seen that for $am_q \gtrsim 0.4$ the residual mass does not remain well controlled. For this reason we restricted the following \emph{M\"obius} Domain Wall fermion simulations to $am_q\leq0.4$ and effectively increased $L_s$ by choosing it as $L_s^\text{M\"obius}=12$ (corresponding to $L_s^\text{Shamir}=24$) as opposed to $L_s^\text{Shamir}=16$ as in figures \ref{fig:kink} and \ref{fig:mres}.

\section{Analysis and Results}
\begin{figure}
\begin{center}
\includegraphics[width=.45\textwidth]{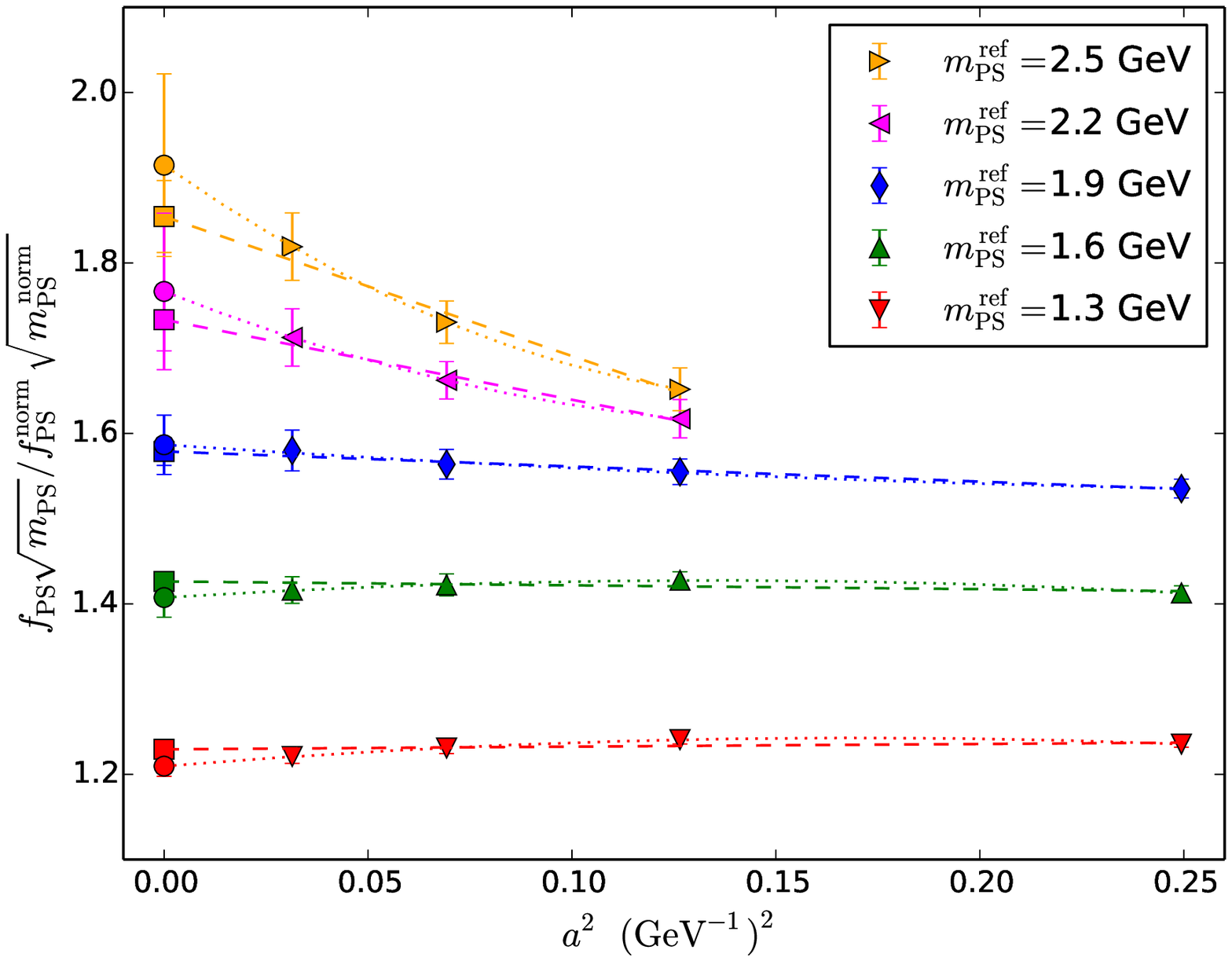}
\includegraphics[width=.45\textwidth]{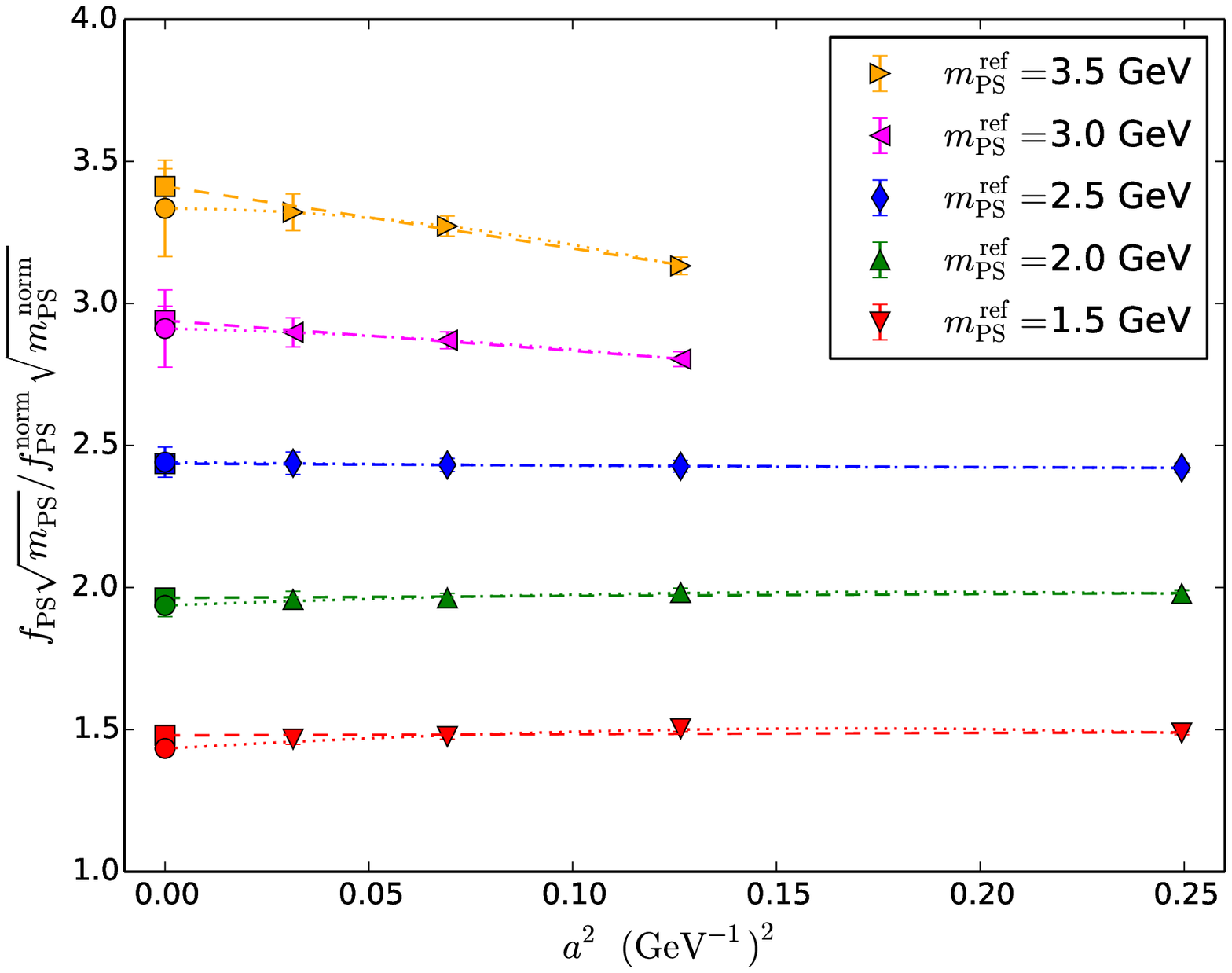}
\end{center}
  \caption{Continuum limit of the ratio of decay constants at different pseudo scalar masses for $D_s$ (left) and $\eta_c$ (right) using M\"obius Domain Wall Fermions. The data is normalised at $m^\mathrm{norm}_\mathrm{PS}=1.0\mathrm{GeV}$. We compare two different polynomial fit ans\"atze, namely linear (dashed lines, square-symbols) and quadratic (dotted lines, round symbols) polynomials in $a^2$. For the heaviest two reference masses the coarsest ensemble does not allow a simulation with $am_q\leq 0.4$, so the continuum limit is taken from the three finer ensembles only.}
\label{fig:cont}
\end{figure}

The physical quantities of which we investigate the continuum scaling are the decay constants of strange-strange ($\eta_s$), heavy-strange ($D_s$-like) and heavy-heavy ($\eta_c$-like) pseudo scalar mesons. Our strategy is as follows:

On each ensemble we simulate two closely spaced strange quark masses and various 'heavy' quark masses with $am_q \leq 0.4$ and compute pseudo scalar masses and matrix elements. $D_s$ and $\eta_s$ masses are interpolated to the \emph{physical} strange quark mass, by matching the $\eta_s$ mass to the published value \cite{eta_s_val}. To be able to take a continuum limit at the same physical point the decay constants for $\eta_c$ and $D_s^\mathrm{phys}$ are now interpolated to common reference masses $m_\mathrm{PS}^\mathrm{ref}$ on each ensemble. Since we are not interested in physical predictions in the quenched approximation we do not calculate renormalisation constants for these quantities, but instead take ratios at a reference mass to cancel the renormalisation constant. This reference mass is chosen as $m_\mathrm{PS}^\mathrm{norm} = 1.0 \mathrm{GeV}$. In sight of further investigations towards the feasibility of $B$-physics and contact to HQET we will consider the expression $f_\mathrm{PS}\sqrt{m_\mathrm{PS}}$ which approaches a constant in the static limit. The resulting continuum limit is presented in figure \ref{fig:cont}. From the two panels in figure \ref{fig:cont} it is clear that the continuum approach with the chosen parameters is very flat and well described by $O(a^2)$ scaling. This is further encouraged by the fact that the residual mass remained well behaved and under control for the simulated data points, confirming the $O(a)$ improvement.

\section{Conclusion and Outlook}
From these results we conclude that M\"obius Domain Wall fermions are a suitable discretisation for charm physics. We expect the qualitative behaviour of our findings to remain unchanged when going beyond the quenched approximation allowing us to dynamically simulate QCD with $2+1f$ whilst keeping discretisation errors under control. This is very encouraging for our ongoing dynamical efforts (first results were presented in J\"uttner's talk at this conference \cite{juettner:Lattice14}).

Finally this quenched data will be used to test the feasibility of doing $B$-physics with Domain Wall fermions, by combining it with results in the static limit as well as using the ratio method \cite{Blossier:2009hg}.

\acknowledgments{The research leading to these results has received funding from the European Research Council under the European Union's Seventh Framework Programme (FP7/2007-2013) / ERC Grant agreement 279757. The authors gratefully acknowledge computing time granted through the STFC funded DiRAC facility (grants ST/K005790/1, ST/K005804/1, ST/K000411/1, ST/H008845/1). PB acknowledges support from STFC grants ST/L000458/1 and ST/J000329/1.}

\end{document}